\documentclass[aps, groupedaddress, prl,showpacs,twocolumn]{revtex4}

\usepackage{graphicx}
\usepackage{dcolumn}
\usepackage{amsmath}
\usepackage{bm}
\usepackage[usenames]{color}
\begin{document}

\title {Quantum complementarity of microcavity polaritons }

\author{S. Savasta$^1$, O. Di Stefano$^1$, V. Savona$^2$, W. Langbein$^3$\footnote{Present address:
Department of Physics and Astronomy, Cardiff University, 5 The Parade, Cardiff CF24 3YB,UK}}
\affiliation{$^1$Dipartimento di Fisica della Materia e Tecnologie Fisiche Avanzate,
Universit\`{a} di Messina, Salita Sperone 31, 98166 Messina, Italy\\
$^2$Institut de Physique Th\'eorique, Ecole Polytechnique F\'ed\'erale de Lausanne (EPFL), CH-1015 Lausanne, Switzerland\\
$^3$Experimentelle Physik IIb, Universit\"{a}t Dortmund, Otto-Hahn-Str.
4, 44221 Dortmund, Germany}

\date{\today}

\begin{abstract}
We present an experiment that probes polariton quantum correlations
by exploiting quantum complementarity. Specifically, we find that
polaritons in two distinct idler-modes interfere if and only if they
share the same signal-mode so that ``which-way'' information cannot
be gathered. The experimental results prove the existence of
polariton pair correlations that store the ``which-way''
information. This interpretation is confirmed by a theoretical
analysis of the measured interference visibility in terms of quantum
Langevin equations.

\end{abstract}

\pacs{71.36.+c, 42.50.Dv, 71.35.Gg, 42.50.Nn}

\maketitle

Quantum complementarity is the essential feature distinguishing
quantum from classical physics \cite{OScully1991}. When two physical
observables are complementary, the precise knowledge of one of them
makes the other unpredictable. The most known manifestation of this
principle is the property of quantum-mechanical entities to behave
either as particles or as waves under different experimental
conditions. The link between quantum correlations, quantum
nonlocality and Bohr's complementarity principle was established in
a series of ``which-way'' experiments
\cite{Zeilinger1995,MandelPRL,MandelReview,delayedchoice}, in which
the underlying idea is the same as in Young's double-slit
experiment. Due to its wave-like nature, a particle can be set up to
travel along a quantum superposition of two different paths,
resulting in an interference pattern. If however a ``which-way''
detector is employed to determine the particle's path, the
particle-like behavior takes over and an interference pattern is no
longer observed. These experiments have brought evidence that the
loss of interference is not necessarily a consequence of the back
action of a measurement process \cite{delayedchoice}. Quantum
complementarity is rather an inherent property of a system, enforced
by quantum correlations \cite{OScully1991}. We investigate this
manifestation of quantum mechanics for cavity polaritons. Polaritons
in semiconductor microcavities are hybrid quasiparticles consisting
of a superposition of cavity photons and two-dimensional collective
electronic excitations (excitons) in an embedded quantum well
\cite{Weisbuch}. Owing to their mutual Coulomb interaction, pump
polaritons generated by a resonant optical excitation can scatter
resonantly into pairs of polaritons (signal and idler)
\cite{Ciuti,LangbeinPRB04,savasta1996,Savasta2003,Houdre,Stevenson,Erland}.
In the low excitation limit, the polariton parametric scattering is
a spontaneous process driven by vacuum-field fluctuations
\cite{savasta1996} whereas, already at moderate excitation
intensity, it displays self-stimulation \cite{Stevenson}.

In this letter we present a ``which-way'' experiment based on the polariton 
parametric process. Our findings demonstrate the pair-correlation of the 
emitted signal-idler polaritons \cite{Ciutisqueez}. This fact, together with the 
possibility of ultrafast optical manipulation and ease of integration of these 
micro-devices, holds promise for applications in quantum information science.

\begin{figure}[!t]
\includegraphics[width=0.55\textwidth]{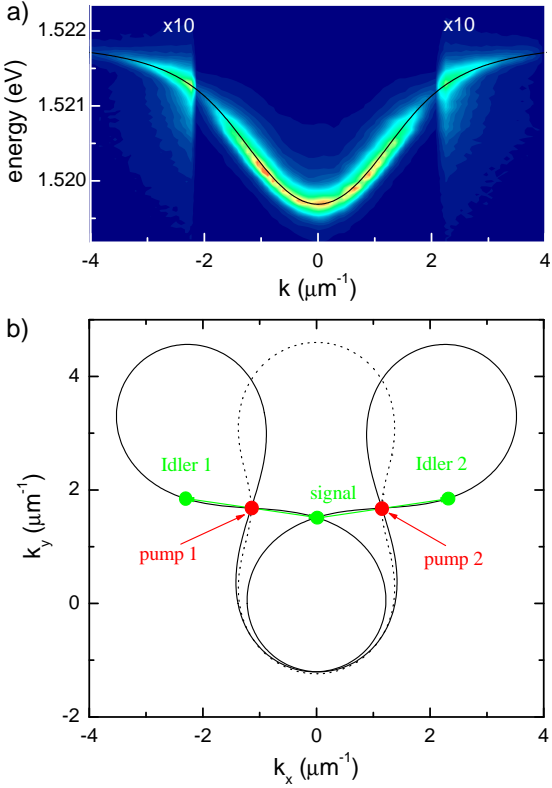}
\caption{(a) Lower polariton photoluminescence intensity (linear
color scale) as function of energy and $|{\bf k}|$. Excitation is in
the upper polariton branch and the detuning between the cavity
resonance and the energy of the heavy-hole exciton is $-0.7$ meV.
The dispersion $E_{\bf k}$ (full line) calculated with a
coupled-oscillator model \cite{LangbeinJPhys04} is superimposed. (b)
${\bf k}$-space plot of the final states fulfilling energy and
momentum conservation in a two-pump parametric process. The two
``eight''-shaped full curves represent the single-pump processes
determined by the conditions $2{\bf k}_{{\rm p}j}={\bf k}_{\rm
s}+{\bf k}_{\rm i}$ and $2E_{{\bf k}_{{\rm p}j}}=E_{{\bf k}_{\rm
s}}+E_{{\bf k}_{\rm i}}$, with $j=1,2$. The dotted line describes
the mixed-pump process defined by ${\bf k}_{\rm p1}+{\bf k}_{\rm
p2}={\bf k}_{\rm s}+{\bf k}_{\rm i}$ and $E_{{\bf k}_{\rm
p1}}+E_{{\bf k}_{\rm p2}}=E_{{\bf k}_{\rm s}}+E_{{\bf k}_{\rm i}}$.
Two parametric processes sharing ${\bf k}_{\rm s}$, giving rise to
mutual idler coherence are indicated in green.}
\label{fig:dispersion}
\end{figure}

Polariton parametric scattering is a $\chi^{(3)}$ process in which
two pump polaritons are scattered into a signal-idler polariton
pair, conserving total energy and momentum. The effective
Hamiltonian describing the parametric polariton process assuming
classical pump-fields is
\begin{equation} \hat{H}=\sum_{\bf k}E_{\bf k}\hat{p}^\dagger_{\bf
k}\hat{p}^{}_{\bf k}+ \sum_{ \stackrel{{\bf k},{\bf
k}^\prime}{\mbox{$\scriptscriptstyle {\bf k}_{\rm s},{\bf k}_{\rm i}$}}
}\left[G({\bf k},{\bf k}^\prime)\hat{p}^\dagger_{{\bf k}_{\rm
s}}\hat{p}^\dagger_{{\bf k}_{\rm i}}+\mathrm{H. c.}\right]\delta_{{\bf k}_{\rm
s}+{\bf k}_{\rm i},{\bf k}+{\bf k}^\prime}\, , \label{hamiltonian}
\end{equation}
where the Bose operators $\hat{p}^\dagger_{\bf k}$ are the polariton creation operators, $E_{\bf
k}$ is the polariton energy (Fig.\,\ref{fig:dispersion}a), and $G({\bf k},{\bf k}^\prime)$ contains
details of the pump fields and the polariton interaction. Due to the assumption of classical pump
fields, this Hamiltonian is formally equivalent to that of a $\chi^{(2)}$ parametric downconversion
\cite{MandelReview,ScullyBook}. Given a pump wave vector ${\bf k}_{\rm p}$, the final states of
parametric processes satisfying energy and momentum conservation are represented by an
``eight''-shaped curve in ${\bf k}$-space \cite{Ciuti,LangbeinPRB04}, depicted in
Fig.\,\ref{fig:dispersion}b. A pair of final states (signal and idler) is defined by the
intersections of the curve with a straight line passing through ${\bf k}_{\rm p}$. We denote as
``idler'' the modes with $k>k_{\rm p}$. The experimental scheme that we devise employs two mutually
coherent pump modes of momenta ${\bf k}_{\rm p1}$ and ${\bf k}_{\rm p2}$ with the classical
amplitudes of the two pump-polariton fields $P_{{\bf k}_{\rm p1}}$ and $P_{{\bf k}_{\rm p2}}$, for
which $G({\bf k},{\bf k}^\prime)=gP_{{\bf k}}P_{{\bf k}^\prime}(\delta_{{\bf k},{\bf k}_{\rm
p1}}+\delta_{{\bf k},{\bf k}_{\rm p2}})(\delta_{{\bf k}^\prime,{\bf k}_{\rm p1}}+\delta_{{\bf
k}^\prime,{\bf k}_{\rm p2}})$. The constant $g$ is the polariton-polariton interaction amplitude,
accounting for both the Coulomb interaction and the Pauli exclusion principle \cite{Ciuti}. Two of
the four products of $\delta$'s represent parametric processes driven by a single pump mode. The
two other terms are mixed-pump processes involving one polariton from each pump mode, whose
energy-momentum conservation defines the ``peanut''-shaped curve illustrated in
Fig.\,\ref{fig:dispersion}b (dashed).  Let us first consider a pair produced by one pump only
($P_{{\bf k}_{\rm p2}} =0$). In the limit of low excitation intensity ($\tau = g |P_{{\bf k}_{\rm
p1}}|^2 t \ll 1$), the time evolution operator applied on the polariton vacuum state
$\left|v\right>$ yields the entangled polariton state $|\Psi \rangle = M\left|v\right>_{\rm s,i} +
\tau |1\rangle_{\rm s}|1\rangle_{\rm i}$, where s and i label a pair of signal and idler modes on
the ``eight'' (i.e. $2{\bf k}_{\rm p1}={\bf k}_{\rm s}+{\bf k}_{\rm i}$), and $M$ is a
normalization constant. This quantum state shows that the parametric process produces signal-idler
pair correlations, i.e. that the scattered  polaritons are created in signal-idler pairs. We now
consider two mutually coherent pump polariton fields of equal amplitudes $P_{{\bf k}_{\rm
p2}}=P_{{\bf k}_{\rm p1}}\mbox{e}^{i\phi}$. For this scheme, pairs of parametric processes sharing
the signal mode are allowed, as schematically depicted in Fig.\,\ref{fig:dispersion}b. Such a pair
of processes involves two idler modes i1 and i2 and one common signal mode s. In the limit of low
excitation intensity, the time evolution operator applied on the polariton vacuum state yields in
this case
\begin{equation}
|\Psi\rangle = M \left|v\right>_{\rm s,i1,i2} + \tau\;
|1\rangle_{\rm s}\left(\, |1\rangle_{\rm i1}|0\rangle_{\rm i2} +
e^{-2i\phi}\, |0\rangle_{\rm i1}|1\rangle_{\rm i2}\right)\, .
\label{lowintensity}
\end{equation}
The resulting polariton population at the signal mode $\langle \Psi
|\hat p_{{\bf k}_{\rm s}}^\dag \hat p^{}_{{\bf k}_{\rm s}} | \Psi
\rangle$ is independent of $\phi$. Interference is absent due to the
orthogonality of the superimposed idler states i1 and i2 in the
parentheses. This quantum superposition stores the ``which-way''
information, each term in the sum representing one possible idler
path. Interference is also absent from either idler-polariton
density. A different result is found for the mutual coherence of the
two idler modes, which is observable in the sum of the two idler
polariton fields. The resulting particle population is $\langle \Psi
|(\hat p_{{\bf k}_{\rm i1}}^\dag +\hat p_{{\bf k}_{\rm i2}}^\dag)
(\hat p^{}_{{\bf k}_{\rm i1}} +\hat p^{}_{{\bf k}_{\rm i2}} )| \Psi
\rangle = 2\langle \Psi |\hat p_{{\bf k}_{\rm i1}}^\dag \hat
p^{}_{{\bf k}_{\rm i1}} | \Psi \rangle \left(1+\cos(2\phi)\right)$,
showing interference because the two idler modes are pair-correlated
with the same signal mode, and thus even by a signal-idler
coincidence measurement, no ``which-way'' information could be
retrieved. We point out that similar conclusions are obtained in the
regime of high excitation intensity by assuming classical signal and
idler fields and random noise terms. In this case, signal and idler
phases are random but their sum-phase is locked to the pump phase,
which again results in a mutual idler coherence if and only if they
share a common signal. As we will see below, a quantitative analysis
of the measured visibility supports the quantum interpretation of
the present experiment.

\begin{figure}[!t]
\includegraphics[width=0.45 \textwidth]{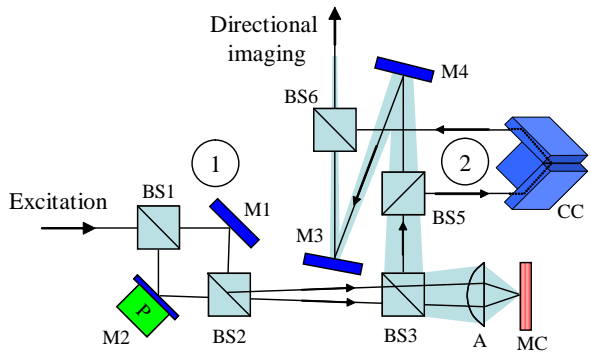}
\caption{Scheme of the optical setup. The two phase coherent pump
pulses are created in a Mach-Zender interferometer
$\bigcirc$\hspace*{-7.5pt}{\scriptsize 1}\hspace*{3pt} consisting of
the mirrors M1, M2 and the beam-splitters BS1, BS2. Their relative
phase is adjustable by a piezoelectric element (P). The pump pulses
are imaged by the aspheric lens A of 0.5 numerical aperture from the
mirrors onto the microcavity. The emission from the microcavity is
collected by the same lens, and directed by BS3 into a second
Mach-Zender $\bigcirc$\hspace*{-7.5pt}{\scriptsize 2}\hspace*{3pt}
which, by using a double mirror (M3,M4) in one arm and a corner cube
mirror (CC) in the other one, produces interference between fields
at ${\bf k}=(k_x,k_y)$ and ${\bf k}^\prime=(-k_x,k_y)$. All
beam-splitters are non-polarizing.} \label{fig:setup}
\end{figure}

Polaritons at a given ${\bf k}$ can be observed by detecting the
photons emitted at the same in-plane momentum \cite{Savona96}, which
carry the polariton amplitude and phase information. In the
experiment we therefore measure the mutual interference of the two
emitted idler fields indicated in Fig.\,\ref{fig:dispersion}b as a
function of the relative pump phase $\phi$. This is accomplished by
detecting two superimposed ${\bf k}$-resolved images of the
polariton emission, one of which is preliminarily mirrored around
the $k_x=0$ axis. The investigated sample \cite{LangbeinPRB04}
consists of a 25\,nm GaAs/Al$_{0.3}$Ga$_{0.7}$As single quantum well
placed in the center of a $\lambda$--cavity with
AlAs/Al$_{0.15}$Ga$_{0.85}$As Bragg reflectors. The wide GaAs
quantum well shows a negligible inhomogeneous broadening of the
fundamental exciton, so that polariton modes have a small broadening
also at large ${\bf k}$. The sample was held at a temperature of
5\,K. The dispersion of the lower polariton branch was determined by
photoluminescence spectra as function of $|\mathbf{k}|$ (see
Fig.\,\ref{fig:dispersion}a). The two mutually coherent pump pulses
of 1\,ps Fourier-limited duration with an adjustable phase delay
$\phi$ are created by splitting the exciting laser pulses (see  Fig.
\ref{fig:setup}), and are synchronously impinging on the microcavity
with $\mathbf{k}_{\rm p1}=(k_{{\rm p}x},k_{{\rm p}y})$ and
$\mathbf{k}_{\rm p2}=(- k_{{\rm p}x},k_{{\rm p}y})$. The pump pulses
are linearly $y$ polarized, and detected is the $x$ polarized
component of the emission. This configuration ensures the detection
of the co-circularly polarized photons which are expected from the
spin-preserving parametric process \cite{Savvidis}, while suppressing the light
elastically scattered by static disorder. The pump intensity is sufficiently low
to rule out parametric processes of higher order, observed instead for higher
pump intensity. The
emitted photon field ${\cal E}_\mathbf{k}$ of the MC is directed
through an interferometer creating the superposition ${\cal
E}_\mathbf{k}+{\cal E}_\mathbf{k'}$, where ${\bf k}=(k_x,k_y)$ and
${\bf k}^\prime=(-k_x,k_y)$.

\begin{figure}[!t]
\includegraphics[width=0.50 \textwidth]{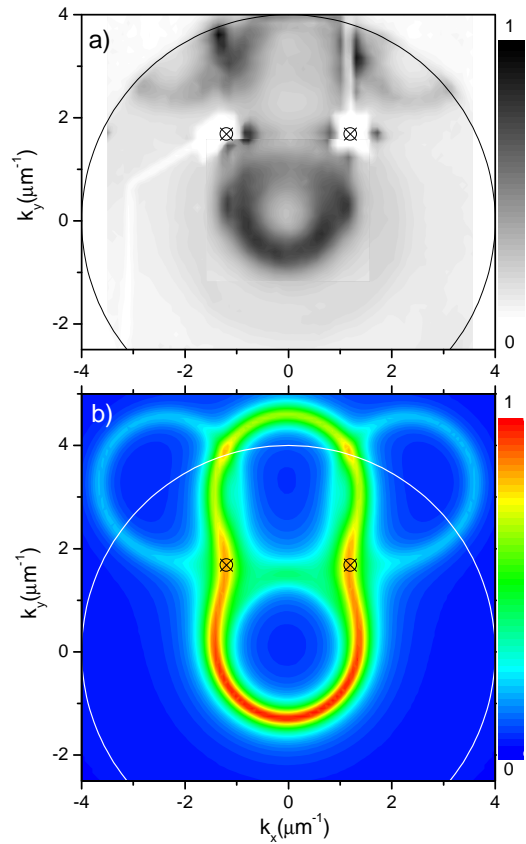}
\caption{\label{fig:dirdens} Polariton density versus $\mathbf{k}$.
The circles indicate the experimentally accessible
$\mathbf{k}$-range, and the crosses the positions of the pump
pulses.  a) Measured for pulsed excitation of
$\approx9\,$nJ/cm$^{2}$, corresponding to an excited polariton
density of $\approx 3\times 10^9$cm$^{-2}$. The white shadows are
masks blocking the pump beams. b) Simulated for cw-excitation using
the experimentally determined polariton dispersion and pump momenta,
and a constant polariton energy linewidth $\Gamma=0.1$ meV.}
\end{figure}

Blocking one of the detection interferometer arms, the intensity
$I_{\bf k} = |{\cal E}_\mathbf{k}|^2$ is measured (see
Fig.\,\ref{fig:dirdens}a), which is proportional to the expectation
value of the polariton number $N_{\bf k}=\langle \Psi|\hat{p}_{\bf
k}^\dagger\hat{p}_{\bf k}|\Psi\rangle$. We have modeled the
parametric polariton emission originating from two pumps in terms of
an extension of the density matrix formalism that was previously
used in the single-pump case \cite{Ciuti,Ciutisqueez}. This model
shares the same physical assumptions as the simple quantum-state
analysis presented here, but allows to obtain the ${\bf k}$-space
dependence of the polariton density. It also accounts for a finite
polariton lifetime, which is mainly due to the photon escape. The
calculated polariton density (Fig.\,\ref{fig:dirdens}b) is in
qualitative agreement with the one deduced from the measured angular
pattern of emission, showing that the polariton parametric process
dominates in the present experimental conditions. The quantitative
deviations are attributed to a deformation of the polariton
dispersion occurring at moderate polariton density \cite{Ciuti} and
to details of the spin-dependent scattering not taken into account.

\begin{figure}[!t]
\includegraphics*[width=0.45\textwidth]{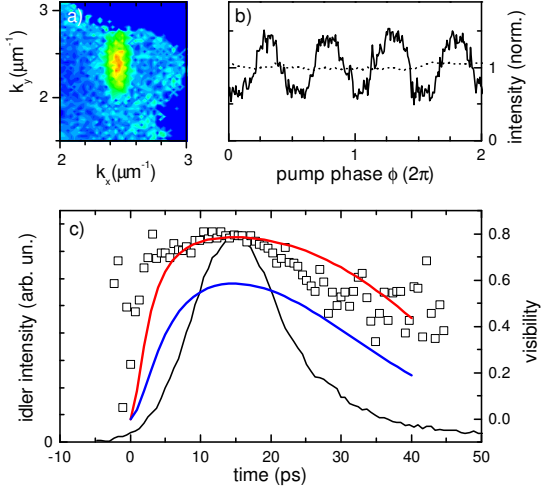}
\caption{Measured interference between parametric emission at
$\mathbf{k}$ and $\mathbf{k}'$. a) $\mathbf{k}$ resolved idler
interference visibility (color scale 0 to 0.5). b) Measured $\tilde
I_\mathbf{k}$ at $\mathbf{k}=(2.4,2.4)\mu$m$^{- 1}$ (full) and at
$\mathbf{k}=(0.0,1.0)\mu$m$^{- 1}$ (dotted) versus pump phase
$\phi$. c) Measured time-resolved idler intensity (full, left scale)
and interference visibility (squares, right scale). In color the
classical (blue) and quantum (red) theoretical predictions of the
visibility.} \label{fig:interference}
\end{figure}

Using both arms of the detection interferometer, the detected
intensity $\tilde I_\mathbf{k}=|{\cal E}_\mathbf{k}+{\cal
E}_{\mathbf{k}^\prime}|^2$, proportional to the polariton number
$\tilde N_{\bf k}=\langle \Psi|(\hat{p}_{\bf
k}^\dagger+\hat{p}_{{\bf k}^\prime}^\dagger)(\hat{p}_{\bf
k}+\hat{p}_{{\bf k}^\prime})|\Psi\rangle$, depends in general on the
relative excitation phase $\phi$ of the pump pulses. To understand
at which $k_x$ the idler polaritons share the same signal mode, and
thus interference is expected, we consider parametric processes
driven by either one of the two pump modes. The pump at
$\mathbf{k}_{\rm p1}$ produces the emission of polariton pairs at
$\mathbf{k}$ and $2\mathbf{k}_{\rm p1}-\mathbf{k}$ respectively.
Equivalently, the pump at $\mathbf{k}_{\rm p2}$ produces pairs of
$\mathbf{k}'$ and $2\mathbf{k}_{\rm p2}-\mathbf{k}'$. By
construction, the modes at $2\mathbf{k}_{\rm p1}-\mathbf{k}$ and
$2\mathbf{k}_{\rm p2}-\mathbf{k}'$ have the same y-component but
opposite x-components $\pm (2k_{{\rm p}x}-k_x)$. Consequently, idler
modes at $\mathbf{k}$ and $\mathbf{k}'$ share the same signal mode
if and only if $k_x = 2k_{{\rm p}x}$. We determine the visibility
from the measurements using $V_\mathbf{k}=\sqrt{2[\langle (\tilde
I_\mathbf{k})^2\rangle_\phi/\langle \tilde
I_\mathbf{k}\rangle_\phi^2- 1]}$, where $\langle\cdots\rangle_\phi$
denotes the average over $\phi$. The visibility for the parametric
process depicted in Fig. \ref{fig:dispersion}b is given in
Fig.\,\ref{fig:interference}a, showing that interference is actually
observed at and only at the expected region in $\mathbf{k}$-space.
The measured interference between the two idler polaritons shows a
$\cos(2\phi)$ dependence (see Fig.\,\ref{fig:interference}b), as
predicted by Eq.\,(\ref{lowintensity}). Additionally, in agreement
with the above analysis, the signal at $k_x = 0$ (dotted) displays
no interference, even though it is created by a superposition of
contributions from both pumps.

As our experiment uses pulsed excitation, the idler interference has
a distinct dynamics during the buildup and the decay of the
parametric scattering \cite{LangbeinPRB04}. In
Fig.\,\ref{fig:interference}c we show the measured interference
visibility as a function of time, together with the idler intensity.
The visibility reaches 0.8 soon after the pump pulse, much earlier
than the maximum in the idler-polariton number. A coincidence
detection of signal-idler pairs could directly assess the quantum
origin of the interference \cite{MandelPRL}, but was not available
for the present work. We therefore need to consider that
interference at high field intensity could be equally accounted for by a classical parametric model
including random-noise driving terms \cite{ScullyBook}. The present measurement was performed at the
onset of the self-stimulated regime, corresponding to an excitation
intensity of $40I_0$ according to Ref. \onlinecite{LangbeinPRB04}.
In order to asses the nature of
the measured interference, we have to compare it with the
quantum and the classical prediction. This is done using the quantum
Langevin model \cite{ScullyBook}, generalized to include two idler
modes and a time-dependent pump. We denote by $N_0(t)$ the
time-dependent signal-polariton population, and by
$N_{1,2}(t)=\langle \Psi|(\hat{p}_{{\bf k}_{i1}}^\dagger \pm
e^{-2i\phi}\hat{p}_{{\bf k}_{i2}}^\dagger)(\hat{p}_{{\bf k}_{i1}} \pm
e^{-2i\phi}\hat{p}_{{\bf k}_{i2}})|\Psi\rangle/2$ the
idler-polariton superpositions of which only $N_{1}$ is influenced
by parametric coupling. The analytical solution reads
\begin{widetext}
\begin{eqnarray}
N_{0,1}(t)&=&e^{-2\gamma t}\left[N_{0,1}(0)\cosh^2 (G(t,0))
+\left(N_{1,0}(0)+1 \right) \sinh^2 (G(t,0))\right]\nonumber\\
&+&2\gamma\int_0^t e^{-2 \gamma (t-t')} \left[
\sinh^2 (G(t,t^\prime)) \left( n_{1,0}+1 \right) + \cosh^2 (G(t,t^\prime))n_{0,1} \right]dt'\, ,\label{langevin1}\\
N_2(t)&=&e^{-2\gamma t}N_2(0)+(1-e^{-2\gamma t})n_2\, .\label{langevin2}
\end{eqnarray}
\end{widetext}
Here $\gamma$ is the polariton lifetime, $n_j$ ($j=0,1,2$)
are the populations of the thermal reservoir acting as a noise
source term, and
$G(t,t^\prime)=\int_t^{t^\prime}\sqrt{2}gP^2_{\mathbf{k}_{p1}}(t^{\prime\prime})dt^{\prime\prime}$.
We adjust the pump amplitude to reproduce the measured variation of the idler intensity as a function of the pump intensity
$(1/I_\mathbf{k})dI_\mathbf{k}/dI_P$, where $I_P=|P_{\mathbf{k}_{p1}}|^2$. Both the noise
terms $n_{0,1}$ and the initial conditions $N_{0,1}(0)$ are
taken equal to 0.2, which reproduces the measured photoluminescence
intensity in a momentum region such that
$|\mathbf{k}|=|\mathbf{k}_{i1}|$ but parametric scattering is
absent. With these values, the estimated peak idler-polariton occupation number
per mode is $N_{\mathbf{k}_{i1}}=N_{\mathbf{k}_{i2}}\sim2$. The calculated visibility is plotted as a red curve in
Fig.\,\ref{fig:interference}c. From the same model, by neglecting
the Bose commutator terms ($N+1 \rightarrow N$), we obtain the
classical prediction (blue curve). The comparison of the two results
with the measurement suggests the quantum origin of the measured
interference.

In conclusion, we have devised and performed an experiment where pair-correlated 
states of electronic excitations (polaritons) in a semiconductor system were 
produced. A quantitative analysis of the measured visibility supports the 
quantum interpretation of the present experiment. This result opens the 
possibility of producing many-particle entangled states of light-matter waves in 
a semiconductor, extending further the perspective of using solid-state 
micro-devices for the implementation of quantum information technology.

We are grateful to A.~Callegari, B.~Deveaud, R.~Girlanda,
A.~Quattropani, P.~Schwendimann, and R.~Zimmermann for discussions
and suggestions. W.L. acknowledges support by U.~Woggon. The sample
was grown by J.Riis~Jensen at the Research Center COM and the Niels
Bohr Institute, Copenhagen University. V. S. acknowledges funding
from the Swiss National Science Foundation through project N.
620-066060.

\end{document}